\documentstyle[epsf,here]{lamuphys}
\begin{document}

\title{
   Hydrodynamic Singularities
    }
\author{
Jens Eggers }
\institute{Universit\"at Gesamthochschule Essen, Fachbereich Physik, \\ 
  45117 Essen, Germany}
\maketitle

\section{Introduction}
The equations of hydrodynamics are nonlinear partial differential equations,
so they include the possibility of forming singularities in finite 
time. This means that hydrodynamic fields become infinite or at least 
non-smooth at points, lines, or even fractal objects.
This mathematical possibility is the price one has to pay
for the enormous conceptual simplification a continuum theory 
brings for the description of a many-particle system. Near singularities,
the microscopic structure re-emerges, as the flow changes over 
arbitrarily small distances. Eventually, the singularity is
cut off by some microscopic length scale such as the distance 
between molecules.

The most fundamental question is whether the microscopic structure
becomes relevant for features of the flow much larger than the 
microscopic ones. If this is the case, the continuum description
is no longer self-consistent, but has to be supplemented by 
microscopic information.
There are some well-known cases where singularities are 
artifacts of neglecting diffusive effects like viscosity 
in the hydrodynamic equations, and there is no more smoothing
on small scales. Examples are the singularities widely believed 
to be produced by  
the three-dimensional Euler equation (\cite{M91}) or those of Hele-Shaw
flows at zero surface tension (\cite{DKZ91}). There does not
seem to be a direct correspondence between the intricate 
structure of spatial singularities produced by these equations 
and real flows. For a realistic description of experiments,
some tiny amount of viscosity or surface tension needs to 
be added. 

The reason hydrodynamic singularities have nevertheless 
attracted considerable attention in recent years
is the observation that certain singularities have direct
physical significance and are not a consequence of 
inadequacies of the equations. In particular, free-surface 
flows exhibit a rich variety of experimentally observable 
singularities, which are responsible for phenomena like 
the breakup of jets, coalescence of drops, and bubble entrainment.

\section{Physical singularities}
We attempt to divide these physical singularities 
into two categories.
\subsection{Dynamical singularities}
These are singularities which are confined to a point in 
time. Usually, they are associated with topological transitions
like the breakup of a piece of fluid into two pieces or the 
joining of two pieces into one. For example, the breakup of
a viscous jet of fluid is driven by surface tension, which 
tries to reduce the surface area by diminishing the radius 
of the jet (\cite{E97}). Inertial forces 
constrain the motion to become more and more localized, 
since smaller and smaller amounts of fluid have to move. 
This causes the jet to break at a point in finite time.
Only the smoothing effect of viscosity prevents infinite 
gradients from occurring before the local radius goes to zero.

As the local radius of a fluid thread becomes smaller and
smaller during pinching, it inevitably reaches a microscopic 
scale $\ell_{micro}$ where the equations cease to be valid.
For thread diameters between 10 and 100 nm, 
short-ranged van der Waals forces come into play, and 
for even smaller diameters the concept of a sharp interface will 
certainly loose its meaning. Moreover, a stability analysis 
(\cite{BSN94}) shows that the pinching thread is very sensitive 
to thermal fluctuations. This makes even threads of micron size 
unstable, and leads to a new structure of nested singularities, 
driven by microscopic fluctuations. 
After the thread has dissolved, new surfaces form 
on either side, whose rapid retraction is again governed by 
the Navier-Stokes equation, but with a new topology. 
It would appear as if the continuation to the new Navier-Stokes problem 
should necessarily include the microscopic length $\ell_{micro}$
at which the thread broke. This is however not the case 
(\cite{E97}), as long as one is looking at scales much larger than 
$\ell_{micro}$. Namely,
the final stages occur on very small spatial and temporal scales,
and do not affect the flow at a finite distance away from breakup.
Thus the outer part of the solution can be used as a boundary 
condition for the new problem after breakup. A closer
analysis reveals that this is sufficient to determine the 
new solution completely. This means that the dynamics very quickly 
``forget'' the microscopic details of breaking, thus making
a consistent hydrodynamic theory of the topological 
transition possible.

\subsection{Persistent singularities}
The other important category of singularities are those which 
exist for a period of time, being either stationary, or moving about 
in space like the classical example of a shock wave. 
At finite viscosity, a shock wave is not a true singularity,
but maintains a finite width determined by the ratio of the viscosity
and the shock strength. However, the width $\delta$ of the shock wave is 
typically of the same order as the mean free path of the gas it is moving 
in. What is important is thus the fact that the solution remains 
consistent as $\delta$ goes to zero. Indeed, the dissipation inside 
the shock remains finite in this limit, so on scales much larger than 
$\delta$ the flow field is the same as if $\delta$ were zero. 

\begin{figure}[H]
\begin{center}
\leavevmode
\epsfsize=0.7 \textwidth
\epsffile{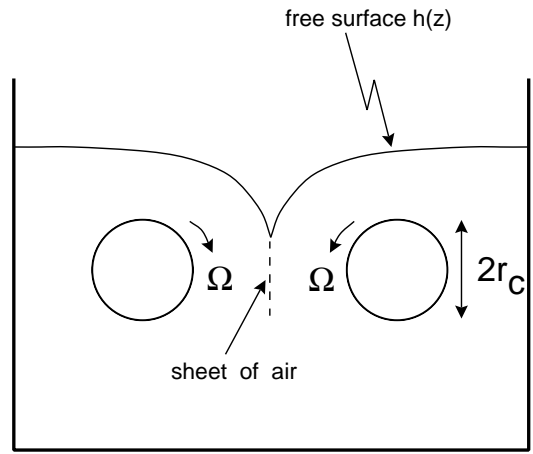}
\end{center}
\caption{Sketch of the two-roller apparatus. At a critical 
capillary number, a pointed cusp appears. At a second, much 
higher capillary number, a sheet emanates from the cusp. }
\label{fig2}
\end{figure}

In the realm of free-surface flows, a beautiful example of 
a stationary singularity has been discovered recently on the surface of 
the viscous flow between two counter-rotating cylinders 
(\cite{JNRR91}). As seen in Fig.\ref{fig2}, two counter-rotating
cylinders are submerged in a container filled with a very 
viscous fluid. The relative strength of viscous forces and
surface tension is measured by the capillary number 
\begin{equation}
\label{Ca}
Ca = \frac{\eta \Omega r_c}{\gamma},
\end{equation}

\noindent
where $\gamma$ is the coefficient of surface tension, $\eta$ 
the viscosity, and $\Omega$ and $r_c$ are the cylinders' 
rotation speed and radius, respectively. At a critical capillary number, 
the surface appears to form a cusp, as indicated in Fig.\ref{fig2}. 
Since the flow remains two-dimensional, this corresponds to a line of 
singularities of the surface. Assuming that it is a true cusp,
the original authors analyzed the flow using a solution due
to \cite{R68}. This local solution leads to a 
logarithmic divergence of the dissipation, and thus cannot be
consistent with continuum theory or a finite driving power. 
It was therefore suggested that the divergence is regularized 
by some microscopic scale.
Since the singularity is very weak, the dissipation 
at the tip would still be small for realistic values of the 
microscopic length. However, this explanation for eliminating 
the logarithmic divergence would mean that there is 
a dependence of the macroscopic flow on a microscopic parameter. 

Faced with this possibility, the problem was reanalyzed by 
\cite{JM92}, who solved the Stokes equation exactly, making the 
simplifying assumption that the two rollers can be represented by a
single dipole. The remarkable result of their calculation 
is that the radius of curvature $R$ at the tip is in fact finite, 
but exponentially small in the capillary number:
\begin{equation}
R = R_0 \exp\left[ -32 \pi Ca\right ] .
\label{R}
\end{equation}
For realistic values of the capillary number, this gives radii 
of curvature far below any physical scale, but is still able to regularize 
the logarithmic singularity. Thus one finds a finite value of the
energy dissipation, making 
the macroscopic flow independent of the microscopic parameters
of the fluid. 

For practical purposes, the theoretical value of 
$R$ is far too small to be realistic. Rather, it most likely 
is the gas above the fluid which will set the value of $R$, and 
this physical effect has been neglected so far. Because of the no-slip 
boundary condition, gas will be forced into the narrow channel 
formed by the cusp. A simple calculation based on lubrication theory 
shows that for $R=0$ this will lead to a diverging pressure at the 
tip of the cusp. Thus the gas will force the channel to widen
to a finite radius, at which the gas pressure is comparable 
to the pressure inside the fluid. 

It is worth noting that the independence from microscopic parameters 
is by no means self-evident. A famous counterexample is that of a
moving contact line, which occurs for example when a circular drop
is allowed to spread on the surface of a table (\cite{dG85}, \cite{BB93}).
Using kinematic arguments alone, one shows that there will be a 
logarithmic singularity of the energy dissipation at the moving 
contact line. There is a vast literature on this problem, 
dealing either with possible mechanisms for a microscopic cutoff,
or with a consistent mathematical description of the resulting 
macroscopic dynamics. The important point to note is that 
continuum mechanics alone cannot resolve the problem in a self-consistent
fashion. It would predict that the spreading of the drop is
stalled, contradicting observation.

\section{Scaling}
The central assumption behind the description of singularities 
is that of locality. Their spatial and temporal scale becomes
arbitrarily small, so that the dynamics should be removed from 
the large-scale features of the flow. However, consistency 
between the singular and the large scale dynamics has to be assured 
by matching conditions between the inner and the outer problems.

A second, closely related assumption is that of self-similarity 
of the singular flow, which seems a natural concept for a class
of problems which lack a typical length-scale.
In the case of time-dependent singularities it means that the 
interface shapes at different times can be mapped on one another 
by an appropriate rescaling of the axes. For example, the 
surface profile of the pinch singularity when a fluid thread 
breaks is (\cite{E97})
\begin{equation}
\label{h}
h(z,t) = \frac{\gamma}{\eta} |t_0 - t| 
\phi\left(\frac{\rho^{1/2}(z-z_0)}{(\eta|t_0-t|)^{1/2}}\right),
\end{equation}
where $t_0$ and $z_0$ are the temporal and spatial position 
where the fluid breaks. Remarkably, the scaling function 
$\phi$ is universal, independent of the type
of fluid or of initial conditions.
For the free surface cusp of Fig.\ref{fig2}, the shape of the 
interface has the scaling form
\begin{equation}
\label{free}
h(z) = h_0 + R^{1/2} f\left(\frac{z}{R^{3/4}}\right),
\end{equation}
where $(0,h_0)$ is the position of the cusps' tip. 

Naturally, it is of particular interest to compute
the scaling exponents.
No general understanding of what selects a particular set
of exponents exists. Usually, local solutions like 
(\ref{h}) or (\ref{free}) are not exact solutions of the equations, 
but only balance certain terms
that are asymptotically dominant. In the case of the pinch
singularity these terms belong to surface tension, viscous, and inertial 
forces. Knowing that, dimensional analysis alone leads to the 
correct power laws. However, there are cases like the pinching of
a very viscous thread (\cite{P95}), where the exponents 
are fixed to irrational values by other consistency 
requirements.

\begin{figure}[H]
\begin{center}
\leavevmode
\epsfsize=0.7 \textwidth
\epsffile{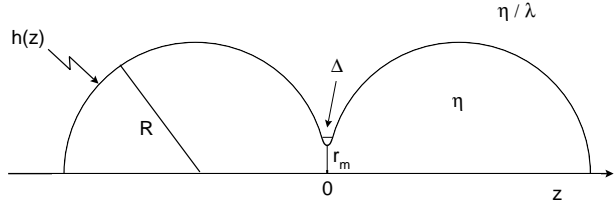}
\end{center}
\caption{Coalescence of two drops of radius $R$ and $R/\delta$.
Asymptotically, the width $w$ of the bridge between the drops 
is much smaller than the minimum radius $r$.
    }
\label{fig3}
\end{figure}

What could possibly keep the local motion from behaving in
a self-similar fashion?
Two examples for broken self-similarity come from the 
{\it coalescence} of two drops. When two drops meet a point, 
surface tension will try to merge them into one drop, 
so the minimum radius $r$ of the bridge between the drops will increase, 
(Fig.\ref{fig3}). As long as the bridge is very small, 
inertial effects can be neglected and the problem is initially governed 
by the Stokes equation. For geometrical reasons the width of the 
bridge $w$ is much smaller than the minimum radius $r$, 
so dimensionally 
\begin{equation}
r(t) = \frac{\gamma}{\eta} t
\label{power}
\end{equation}
is the only possible power law time dependence of the radius. However, 
a closer analysis reveals (\cite{ELS97}) that (\ref{power})
has to be corrected by a logarithmic factor $\log(r / R)$,
where $R$ is the initial radius of a drop. The reason for
this change of the time dependence lies
in the long-ranged character of the Stokes equation, which couples very 
disparate length scales. Hence the width of the bridge cannot 
be neglected, but enters as a logarithmic factor 
$\log(w/r) \sim \log(r/R)$. 

Another mechanism for broken self-similarity is observed for the 
same problem, but for much smaller initial size of the drops. 
In that case the surface-tension-driven motion will first occur 
on the surface alone, rather than being able to drive a flow in the 
interior. The equation of motion for this surface diffusion was first 
given by \cite{M65}. Based on simple scaling arguments, \cite{M65}
gave 
\begin{equation}
\label{1/6}
r(t) \sim t^{1/6}
\end{equation}
as the evolution of the radius, but this result could not be corroborated 
by his own numerical simulations. To understand this failure, one needs 
to take a closer look at the dynamics near the rising bridge
(\cite{E97b}).
In contrast to the viscous flow problem, 
as the gap between the two spheres fills, a bulge 
of material forms just above the minimum. Eventually, it grows enough 
for both sides to touch, forming a void inside the 
material. Of course, at the point of touching a new singularity 
occurs and the topology of the problem has changed. The naive 
assumption of a single continuous evolution, underlying 
(\ref{1/6}), is thus incorrect. Self-similarity 
can at best exist in a discrete sense.

\section{Birth of new structures}
If one drives the two-roller apparatus of Fig.\ref{fig2}
much harder than necessary for the formation of a cusp,
a second transition occurs, 
above which a thin sheet of air emanates from the cusp,
and is drawn continuously into the fluid. This sheet 
is stable in time, but undergoes a three-dimensional
instability at its lower end, where it decays into a curtain of tiny droplets. 
This provides a general mechanism for the entrainment of bubbles
into a flow.
The existence of the sheet was pointed out by \cite{M94}, and confirmed 
in a series of qualitative experiments (\cite{ES96}), using a
silicone oil 10000 times as viscous as water. At sufficiently
high driving, the air forced into the cusp 
experiences a strong enough downward pull
to form a stable sheet. A preliminary 
calculation suggests that the thickness 
of the sheet scales like 
\begin{equation}
\label{delta}
\delta \approx \left(\frac{\eta_{air}}{\eta_{fluid}}\right)^{1/2} r_c 
\quad.
\end{equation} 
This prediction is based on the assumption that there is a return flow 
in the sheet, which produces very high gradients. So when the sheet becomes 
very thin the inner flow is able to balance the high sheer stresses
produced by the viscous flow. Both surface tension and gravity are not
taken into account.

\begin{figure}[H]
\begin{center}
\leavevmode
\epsfsize=0.9 \textwidth
\epsffile{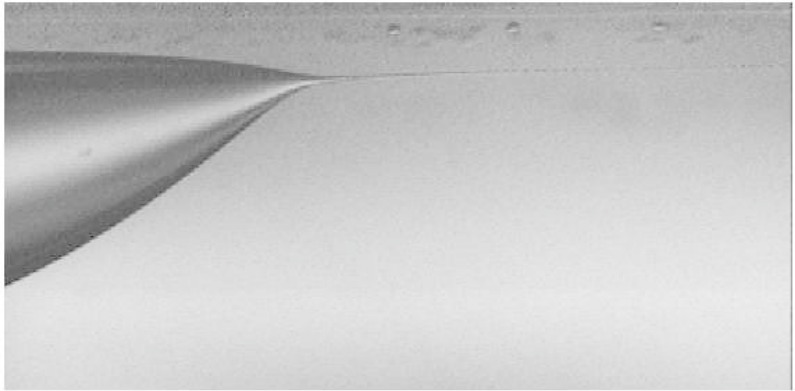}
\end{center}
\caption{Tip-streaming in a Couette device, showing a drop of water 
in silicone oil 1000 times as viscous. To initiate tip-streaming,
200 ppm of surfactant has been added. The image height is 0.5 mm.
Photograph courtesy H. Leonhard.
   }
\label{fig5}
\end{figure}

Similar phenomena have been observed for a variety of 
other stationary singularities. An example, known as 
``tip-streaming'' (\cite{T34},\cite{D93}), is shown in Fig.\ref{fig5}.
One sees a drop of water in a shear flow of a very viscous fluid, which 
produces a cone-shaped singularity at both ends of the drop.
Under circumstances 
that are not well understood, a jet emanates from the tip, 
and eventually decays into drops due to the Rayleigh capillary instability. 
This is the precise analogue of the sheet in the two-roller
apparatus, but the dimension of the singularity and of the resulting 
structure are lowered by one. 
A second example of a zero-dimensional singularity giving rise to 
a one-dimensional structure is that of a dielectric drop in 
a strong electric field, where a local cone-shaped solution 
exists (\cite{T64}).
This ``Taylor cone'' is never stable,
but either oscillates between a rounded and a pointed state, or
stabilizes itself by ejecting a jet from its tip. Again, little is known about 
the conditions under which the jet forms,but the striking similarities
between different systems suggest a unifying explanation for
the emergence of these structures.

\end{document}